\documentclass{sig-alternate}
\pdfoutput=1
\usepackage{latexsym}
\usepackage{amsmath}
\usepackage{amssymb}
\usepackage{graphicx}
\usepackage{algorithm}
\usepackage{algorithmic}
\usepackage{verbatim}
\usepackage{url}

\newcommand{\mathsym}[1]{{}}

\graphicspath{{.}}

\begin{document}

\title{Putting Recommendations on the Map -- Visualizing Clusters and Relations}

\author{
Emden Gansner, Yifan Hu, Stephen Kobourov, Chris Volinsky\\
        AT\&T Labs -- Research, 180 Park Ave, Florham Park, NJ 07932\\
\{erg, yifanhu,skobourov,volinsky\}@research.att.com
}

\maketitle
\pagestyle{plain}
\begin{abstract}

For users, recommendations can sometimes seem odd or counterintuitive.
Visualizing recommendations can remove some of this mystery, showing how
a recommendation is grouped with other choices. A drawing can also lead
a user's eye to other options. Traditional 2D-embeddings of points can be
used to create a basic layout, but these methods, by themselves, do not 
illustrate clusters and neighborhoods very well.
In this paper, we propose the use of geographic maps to enhance the
definition of clusters and neighborhoods, and consider the
effectiveness of this approach in visualizing similarities and recommendations
arising from TV shows and music selections.
All the maps referenced in this paper can be found
in \url{www.research.att.com/~volinsky/maps}. 

\end{abstract}

\section{Introduction}

Information visualization techniques are often essential in helping to make
sense out of large data sets. High-dimensional data can be visualized as a
collection of points in 2-dimensional space using principal component
analysis~\cite{jolliffe-pca}, multidimensional scaling~\cite{MDS},
force directed algorithms~\cite{Fruchterman_Reingold_1991},
or non-linear dimensionality reduction like
LLE/Isomap~\cite{lle,isomap}. These embedding algorithms tend to put similar
items next to each other. Visual examination often suffices to
identify the presence of clusters. Sometimes, however, the clusters
are not as easy to see and additional visual clues are needed to
highlight them. One possibility is to use cluster analysis algorithms,
such as $k$-means or hierarchical clustering
algorithms~\cite{Johnson_1967, Lloyd_1982} to explicitly define
clusters. 
The points and labels can then be colored based on
the clustering. 
While in small examples it is possible to convey the  cluster information just with the use of colors and proximity, this becomes difficult to do with large data. Common problems include dense clusters where labels overlap each other and clusters that lack clearly defined boundaries.

\begin{figure*}[t]
\begin{center}
\includegraphics*[width=18cm, angle = 0]{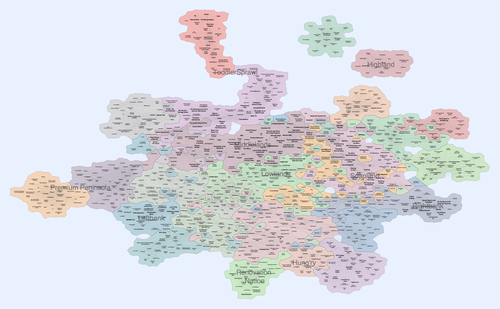}
\vspace{-.3cm}
\caption{\small\sf TVLand: a map based view of the TV similarity graph, depicting the relationship among the 1000 most watched TV shows.  Each show is linked to 10 most similar shows and map clusters, or {\em countries} are represented by different colors. Zoom into the graph to read the labels and see more structure or view the images at
\protect\url{www.research.att.com/~volinsky/maps}. \label{TV_land}}
\end{center}
\end{figure*}

In this paper we propose the use of maps as a way to achieve this
explicit visual definition of clusters. There are several reasons that
such a representation can be more useful. First, by explicitly
defining the boundary of the clusters and coloring the regions, we make
the clustering information clear. Second, as most
dimensionality reduction techniques lead to a 2-dimensional
positioning of the data points, a map is a natural
generalization. Finally, while graphs, charts, and tables often
require considerable effort to comprehend, a map representation is
more intuitive, as most people are very familiar with maps and even
enjoy carefully examining maps. Applying this approach to a data set containing show-show similarities between 1000 TV programs results in the map in Figure~\ref{TV_land}, which conveys clustering information much better.\footnote{
This paper contains zoomable high resolution images; all the images are also available at \url{www.research.att.com/~volinsky/maps}.}

We have considered several different approaches for generating maps,
depending on (1) how we obtain the 2D positions for the
data points, (2) how we cluster the data points, and (3) how we
represent the resulting layout and clustering as a map. Depending on
the application, some choices are more suitable than others. Consider
the case when the data points are TV programs and the goal is to
visualize ``TVLand''. In
this case, it is highly desirable to obtain a map in which similar
programs are placed close to each other and, even better,
grouped in {\em countries} such as {\em Sportsitania},
{\em Newsistan}, {\em ToddlerSprawl}, etc.  Once the data is represented as countries on a map, recommendations can be visualized statically and interactively.  

%

The map metaphor becomes more powerful as a user becomes familiar with
the canonical map layout.  Humans are comfortable with map-related concepts: 
items within a country are similar to each other; 
areas separated by a mountain range are difficult to connect; 
islands might have atypical qualities, etc. We hope through this work to extend 
the familiar map metaphor to the world of recommendations.  Thus, by putting items like TV shows on a map, we can borrow map-related cognitive concepts.
%
In a user-driven mode, a personalized ``heat map''  is generated, where regions of low interest are colored with cool colors and regions containing highly recommended shows are colored with a hot color. These maps are generated dynamically based on user preference and the available TV shows at this moment in time. The user can interactively explore the map to find related shows in the same region, or in neighboring regions. Building ``roads'' between regions, visiting ``islands'' and ``traveling cross-country'' are all metaphors that 
can have meaning in the recommendation space.  In a user-passive mode, recommendations are highlighted on the map. These maps help the user (as well the designers of the recommender system) understand why certain shows are recommended. By 
highlighting both the recommended shows and shows the user has watched on the same map, 
the user understands the proximity of the recommended items and
watched items, thus the rationale for the recommendation.


\section{The mapping algorithm}

\begin{figure*}[t]
\begin{center}
\includegraphics*[width=3cm, angle = 0]{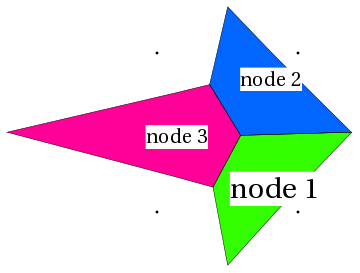}
\hspace{1cm}
\includegraphics*[width=3cm, angle = 0]{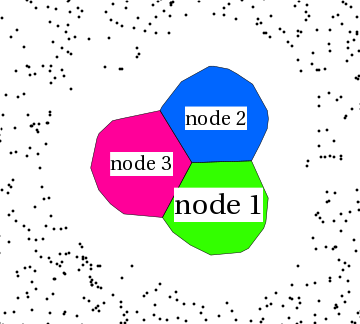}
\hspace{1cm}
\includegraphics*[width=4cm, angle = 0]{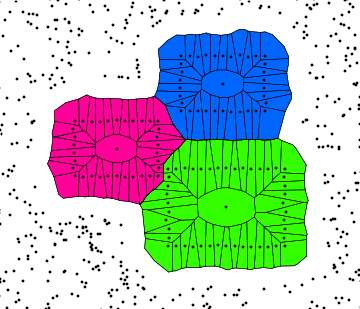}
\hspace{1cm}
\includegraphics*[width=3cm, angle = 0]{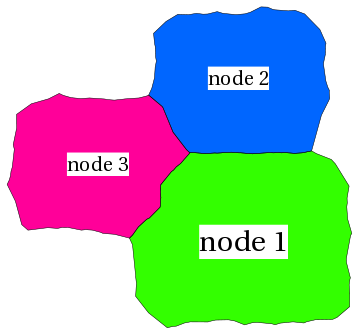}\\
(a) \hspace{4cm} (b) \hspace{4cm} (c) \hspace{4cm} (d)
\caption{\small\sf (a) Map from a Voronoi diagram of vertices and corners of the bounding box. (b) Better construction of outer boundaries through placement of random points. 
(c) Voronoi diagram with additional points  around the bounding boxes of the labels. (d) the final map.
\label{example}}
\end{center}
\end{figure*}

The input to our algorithm is a relational data set from which we
extract a graph $G=(V,E)$ with $|V|=n$ and $|E|=m$. The set of vertices
$V$ corresponds to the objects in the data, e.g., TV programs, and the
set of edges $E$ corresponds to the relationship between pairs of
objects, e.g., the similarity between a pair of shows. In its full
generality, the graph is vertex-weighted and edge-weighted, with vertex
weights corresponding to some notion of the importance of a vertex and
edge weights corresponding to some notion of the distance between a
pair of vertices. In the case of TV programs, the importance of a
vertex can be determined by the popularity of a show as derived from the total number of viewers or by the number of hours watched by one particular user. The weight of an edge connecting two programs can be determined by textual comparison of their program description using a content based model (e.g., \cite{LDA}), or by analyzing viewership using a collaborative filtering based model~\cite{cf}.


In the next step, the graph is embedded in the plane. 
Possible embedding algorithms include
principal component analysis~\cite{jolliffe-pca}, 
multidimensional scaling (MDS)~\cite{MDS},
force directed algorithm~\cite{Fruchterman_Reingold_1991},
or non-linear dimensionality reduction like LLE/Isomap~\cite{lle,isomap}.
A cluster analysis is then performed in order to group vertices into clusters. It is important to match
the clustering algorithm to the embedding algorithm. For example, a geometric
clustering algorithm such as $k$-means~\cite{Lloyd_1982} may be suitable for
an embedding derived from MDS, as the latter tends to place similar vertices in the same geometric 
region with good separation between clusters. On the other hand, 
with an embedding derived from a force directed algorithm~\cite{Fruchterman_Reingold_1991},
a modularity based clustering algorithm~\cite{Newman_2006} is appropriate. The two algorithms are
strongly related, as pointed out in  the recent findings by Noack {\em et al.}~\cite{Noack_2009}, and therefore we can expect vertices that are in the same cluster
to also be geometrically close to each other in the embedding.

The final map is obtained in the last step. Using the layout information, a Voronoi diagram of the vertices is created. 
A naive approach would be to form the Voronoi diagram
of the vertices, together with fours points on the four corners of the bounding box; see Fig.~\ref{example}(a). 
This would result in aesthetically unappealing maps with unnatural outer boundaries and sharp corners. 
A more natural appearance can be obtained by placing some random points that are sufficiently
far away from the set of real vertices, which leads to more rounded boundaries. The randomness of
the points on the outskirt also gives rise to some randomness of the outer boundaries, 
thus making them more realistic and natural; see Fig.~\ref{example}(b).


In most of our maps {\em countries} have different desirable sizes (e.g., in TVLand the size is determined by a show's popularity). We accomplish this by associating larger {\em countries} with larger label sizes.
To make areas proportional to the label size, we first generate points along the bounding boxes of the labels; see Fig.~\ref{example}(c). To make the inner boundaries more realistic, we perturb these points randomly
instead of running strictly along the rectangle bounding boxes. 
Voronoi cells that belong to the same vertex are colored in the same color, and cells that correspond to the random points on
the outskirts are not shown. Cells of the same color are then merged to give the final map; see Fig.~\ref{example}(d).
Note that instead of the bounding boxes of labels, we could use any 2D shapes, e.g., the outlines of  
real countries, in order to obtain a desired look.

When mapping vertices that contain cluster information, in addition to merging cells that belong to the same vertex, we also merge cells that belong to the same cluster, thus forming regions of complicated shapes, with multiple vertices and labels in each region. At this point we can add more geographic components to strengthen the map metaphor.  For instance, in places where there is significant space between vertices in neighboring clusters, we can add lakes, rivers, or mountain ranges, in order to indicate the distance.  These structures can all be formed by similar insertion of random points in places where vertices are far away from each other. 

\section{Visualizing TVLand \& MusicLand}
Many recommender systems rely on knowledge about how items are
related to each other through a similarity measure. Similarity
can be explicitly calculated as in the neighborhood model
\cite{Koren_2008}, or implicitly used as in the factorization model
\cite{Hu_Koren_Volinsky_2008}. Recommendations are then made by
selecting items that are most similar to those that the user has
already sampled. While this provides a useful service, it is
not obvious to the user (and often the designer of the recommender system) why the recommender makes these specific
recommendations. A visualization of all the involved items in the same map would reveal a lot of information
about how items, and clusters of items, relate to each other. It would
serve as the main map from which recommendations can be presented to
the user by zooming in to the area of interest. We apply our mapping algorithm to create the lands of TV and music and illustrate this approach.

\subsection{The land of TV\label{sec_tvland}}

Figure~\ref{TV_land} shows a visualization of the TVLand using our
proposed technique. Our data comes from a digital TV service
with over a million set top boxes\footnote{All the data was collected in
accordance with appropriate end user agreements and privacy policies.
The analysis was done with data that was aggregated and fully
anonymized.  No personally identifiable information was collected in
connection with this research.},
from which we compute the show-show similarity matrix, based on
viewership information. 
There are many ways to compute similarity, but here
we use the approach based on the factorization model, as described by
Hu {\em et al.}~\cite{Hu_Koren_Volinsky_2008}.  The full similarity
matrix is dense and very large (tens of thousands of rows and columns
and millions of non-zero values). For the maps in this paper, we
consider only the top 1000 most popular shows.  For each show, we take
the top 10 most similar shows, which gives us a sparse matrix. The
graph represented by this matrix is embedded using a scalable force
directed algorithm~\cite{Hu_2005}. The font size of each label is proportional
to the popularity of the show and label overlaps are removed in
a post-processing step~\cite{Gansner_Hu_2008}. Clusters are then computed using a modularity
based clustering algorithm~\cite{Newman_2006}. 
The map is colored using a
standard scheme from ColorBrewer~\cite{ColorBrewer}, with additional blending to obtain more colors.

\begin{figure}[h]
\begin{center}
\includegraphics*[width=8cm, angle =0]{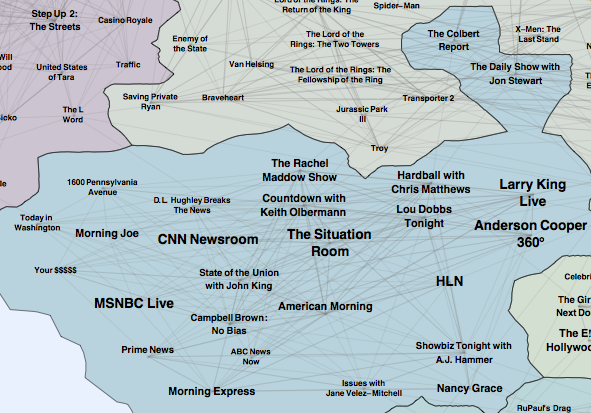}
\vspace{-.3cm}
\caption{\small\sf Leftbank of Newsistan\label{lnewsistan}}
\end{center}
\end{figure}

There are high-level patterns that emerge upon a close inspection of Figure~\ref{TV_land}.  Two cardinal directions seem to be NW-SE and SW-NE. Along the first axis, the NW corner shows are mostly
kid-oriented ({\em Wow Wow Wubbzy} and {\em Hannah Montana}) whereas the SE corner shows seem to cater 
to more mature audiences ({\em Good Day at 7am} and {\em The O'Reilly Factor}). 
Along the second axis, the SW corner has a high concentration of entertainment and fashion shows 
such as {\em What Not to Wear} and {\em E! True Hollywood Show}, whereas in the NE corner, sports and news shows cater to more male audiences.   The bigger fonts of the popular shows is evident in the northern part of the country and in parts of the SW.  

Two islands appear off the NE coast, both due to channel-based clustering. TV viewers often stay on the same channel for extended time intervals, creating strong connections between shows that are not necessarily thematically connected. For example, the {\em Highland Island} contains many popular PBS shows, from {\em Sesame Street} and {\em Barney and Friends} to {\em Charlie Rose} and {\em Antiques Roadshow}. A similar inland cluster directly below 
contains a variety of popular Univision programs, almost all in Spanish. 

The TVLand map contains many descriptive regions.  In the following subsections we take a 
closer look at some of the more interesting ones, seen in Figure~\ref{TV_land} with gray labels.


\begin{figure}[th]
\begin{center}
\includegraphics*[width=8cm, angle=0]{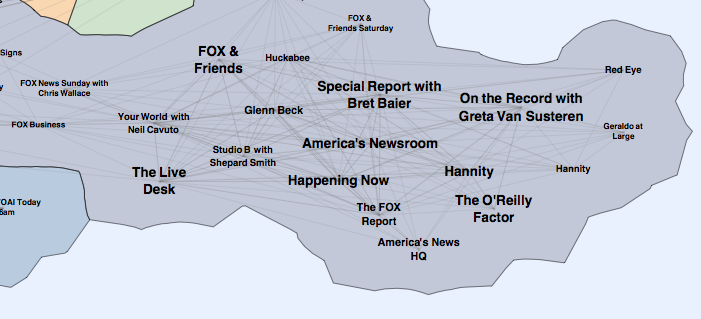}
\vspace{-.3cm}
\caption{\small\sf Rightbank of Newsistan\label{rnewsistan}}
\end{center}
\end{figure}

\paragraph{Newsistan}
Instead of one monolithic cluster of news shows, our map shows several distinct sub-clusters. In the SW of the continent is the compact and well interconnected {\em Leftbank Newsistan} anchored by classic CNN shows {\em Newsroom}, {\em The Situation Room} and {\em Larry King} along with MSNBC newcomers {\em Countdown with Keith Olbermann} and {\em The Rachel Maddow Show} and the off-kilter Comedy Central shows {\em Daily Show with Jon Stewart} and {\em Colbert Report}; see Fig.~\ref{lnewsistan}. Diametrically opposite on the East is the similarly compact and well interconnected {\em Rightbank Newsistan} anchored by {\em Fox and Friends}, {\em Hannity}, and {\em The O'Reilly Factor}; see Fig.~\ref{rnewsistan}. Above are a couple of clusters of local news and below is yet another  news-cluster, mostly made up of morning news shows. It is worth noting that the seemingly meaningful left-right placement of the two distinct news clusters was coincidental. However, the diametrically opposing placement of these two clusters is meaningful as, although they both contain news-related shows, 
there are very few viewers who regularly watch shows in both clusters.

\begin{figure}[h]
\begin{center}
\includegraphics*[width=8cm, angle = 0]{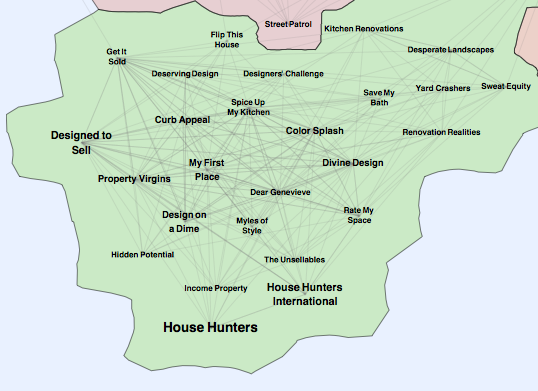}
\vspace{-.3cm}
\caption{\small\sf Renovation Nation\label{renovation}}
\end{center}
\end{figure}
%

\begin{figure}[h]
\begin{center}
\includegraphics*[width=8cm, angle = 0]{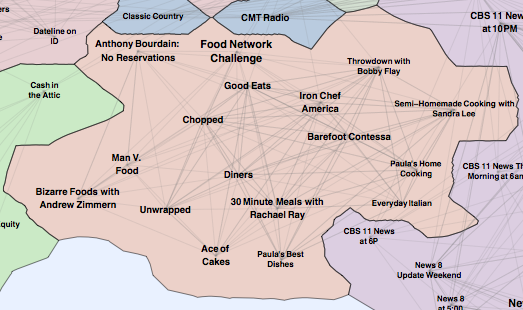}
\vspace{-.3cm}
\caption{\small\sf Hung'ry\label{hungry}}
\end{center}
\end{figure}

\paragraph{Renovation Nation and Hung'ry}
The southernmost tip of TVLand contains a gaggle of HGTV home improvement and real-estate shows we call {\em Renovation Nation} such as {\em Designed to Sell}, {\em House Hunters}, {\em Get it Sold}, and {\em Property Virgins}; see Fig.~\ref{renovation}. Right next door is the {\em Hung'ry} cluster 
with shows {\em Man V. Food} and {\em Iron Chef}; see Fig.~\ref{hungry}.  Even within these regions, 
we can see sub-regions.  The NW corner of {\em Renovation Nation} focuses on selling homes, while other parts of the country deal with renovations or design issues.  The shows featuring celebrity female chefs cluster in the SE bloc of {\em Hung'ry}.

\begin{figure}[htb]
\begin{center}
\includegraphics*[width=6cm, angle = 0]{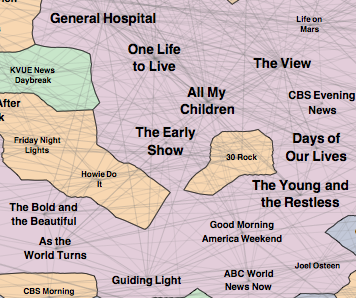}
\vspace{-.3cm}
\caption{\small\sf Soapland\label{soap}}
\end{center}
\end{figure}

\begin{figure*}[t]
\begin{center}
\vspace{-.3cm}
\includegraphics*[width=18cm, angle = 0]{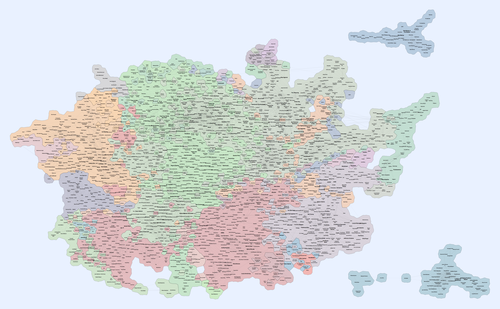}
\vspace{-.3cm}
\caption{\small\sf Map of MusicLand \label{lastfm_all}}
\end{center}
\end{figure*}

\paragraph{ Middlelands, Lowlands and Soapland}
Many of TV's most popular shows are in the middle of TVLand. 
Classic sitcoms such as {\em Seinfeld} and {\em Frasier} and contemporary shows such as {\em House}
and {\em Two and a Half Men} share the same space with crime shows like {\em Law and Order} and {\em CSI}, talk shows such as {\em Late Show with David Letterman} and {\em Tonight Show with Jay Leno}, entertainment shows such as {\em TMZ} and {\em Access Hollywood}, talent shows such as {\em American Idol}  and sports programming such as {\em NASCAR Sprint Cup} and {\em NBA Basketball}.   Directly below the Middlelands of popular shows lie the Lowlands of reality-based court shows {\em Judge Judy}, {\em Judge Joe Brown} and {\em The People's Court}.   In the same cluster we find notorious tabloid talk shows like {\em Jerry Springer} and {\em Maury}.
Next to the Lowlands is the domain of daytime soap operas with the daytime staples of each network represented; see Fig.~\ref{soap}.
 ABC's {\em General Hospital}, {\em One Life to Live}, {\em All My Children} share the same space with CBS's {\em Young and the Restless}, {\em Guiding Light}, {\em As the World Turns} and {\em The Bold and the Beautiful}. NBC's contribution to this cluster is {\em Days of Our Lives}.

\paragraph{ToddlerSprawl}
The NW corner of TVLand consists almost exclusively of kid-oriented shows.
The northernmost point has a prominent peninsula dominated by Nickelodeon powerhouses like {\em Dora}, {\em Ni Hao} and {\em Go}.  This connects to the mainland via a Disney cluster with crowd-pleasers {\em Hannah Monana} and {\em The Suite Life of Zack and Cody}. ToddlerSprawl continues on the north shore of the mainland with cross-over hit shows like {\em SpongeBob SquarePants}. At the NW tip of ToddlerSprawl are Cartoon Network's {\em Johhny Test} and mainstays {\em Tom and Jerry}, {\em Flintstones} and {\em Scooby Doo}.   Conspicuously absent are {\em Sesame Street} and {\em Barney and Friends} which were clustered separately with other PBS shows.

\paragraph{Premium Peninsula}
The western coast of TVLand consists of two monolithic clusters corresponding to the premium channels HBO and Showtime.
The HBO peninsula is represented by popular original series such as {\em Real Time with Bill Maher} and {\em Big Love} and several blockbuster movies. Similarly, its neighbor, the inland Showtime cluster has some original series such as {\em The L Word} and dozens of popular movies. Most of the other movies in TVLand can be found in the immediate vicinity of HBO and Showtime.

\subsection{The land of music}\label{sec_lastfm}

\begin{figure*}[thb]
\begin{center}
\vspace{-.3cm}
\includegraphics*[width=17.5cm, angle = 0]{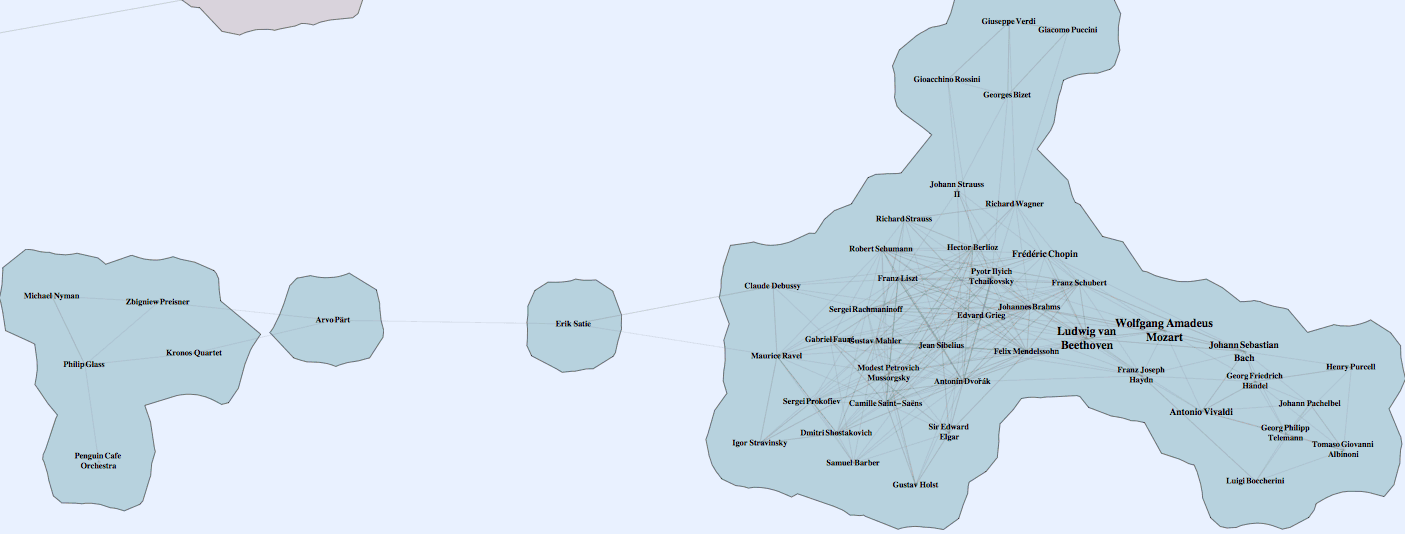}
\vspace{-.3cm}
\caption{\small\sf Classical archipelago\label{classical}}
\end{center}
\end{figure*}

We now turn to the visualization of the land of music. We collected data from a web crawl of the {\tt last.fm} website. As an Internet radio and music community website, it has over 30 million users. 
Using a music recommender system, {\tt last.fm} recommends music based on user profiles. Over several years the recommender system has collected information about how one musician is related to another in terms of how many listeners of one also enjoy the other. For each musician, the last.fm website lists the top 250 related
musicians. For example, Beethoven is considered to have ``super
similarity'' to Mozart, Bach, Brahms etc., ``very high similarity'' to
Mendelssohn, Schumann, Vivaldi, etc.~\footnote{Note that we do not know last.fm's formal definition 
of ``super similarity'', ``very high similarity'', or ``high similarity''.} The website also provides the
number of listeners of each musician. In April 2009, we crawled the
website by starting with Beethoven, and the top 20 musicians most similar to
Beethoven, provided that each has at least 100,000 listeners. We then found
the top 20 most similar musicians to each of those with at least 100,000 listeners and proceed recursively. 
Our crawl yielded a graph with 2782 musicians, with edge weights 
the similarity between musicians. We further prune this graph by only taking edges that have ``super
similarity''
and after taking the non-trivial components we end up with 2588 vertices. We then lay out the graph, cluster the vertices, and generate the MusicLand 
map; see Fig.~\ref{lastfm_all}.

\paragraph{The Mainland}
The vast majority of musicians and bands is located in the continent. While it is not as easy to spot major trends along the main axes, many of the clusters are well-defined and neighboring clusters make sense from a musical point of view. 
The {\em Rocky Coast} cluster on the east shore begins with {\em Eric Clapton} and the {\em Beach Boys} in the south, goes through the {\em Rolling Stones} and {\em Neil Young} in the middle before reaching the {\em Metallands} cluster of {\em AC/DC} and {\em Black Sabbath} in the north. In close proximity to the southern edge of the Rock cluster is a small but major {\em Beatles cluster}. To the north is the {\em Grungelands} cluster with {\em Nirvana, Pearl Jam}, and {\em Soundgarden}.

On the west coast of the main continent we find the {\em Electrolands} cluster of lounge and acid jazz, anchored by {\em Thievery Corporation}, {\em Morcheeba}, and {\em Massive Attack}. 
To the north are avant-garde electronic bands like {\em Autechre, Aphex Twin} and {\em Boards of Canada}. 
In the extreme west are electronic classics {\em Tangerine Dream, Vangelis}, and {\em Jean-Michelle Jarre}. To the south is a compact and well-defined cluster of dance music represented by {\em BT, Paul Oakenfold}, and {\em ATB}. 

In the center of the map, there is a large number of indy bands, the {\em Indyana} cluster, dominated by {\em The Decemberists, The Shins}, and {\em Wilco}.  Nearby is a well-defined concentration of female singer-songwriters: {\em Alanis Morissette, Tory Amos, Sarah McLachlan}, and {\em Tracy Chapman}. 

\paragraph{The Islands} There are two notable islands regions in MusicLand: in the NE is Reggae island, while the chain of islands off {\em Rocky Coast} in the NW make up the {\em Classical Archipelago}.
It is easy to find some general patterns in the layout of {\em Classical archipelago}; see Fig.~\ref{classical}. Two cardinal directions are W-E and N-S. Along the first axis, the west corner contains modern composers such as {\em Ravel, Satie}, and {\em P{\"a}rt}, while the east corner contains 17th century composers such as {\em Bach, Handel,} and {\em Albinoni}. 
Along the second axis, the north corner has a high concentration of opera composers such as {\em Verdi, Rossini}, and  {\em Puccini,} 
whereas the south corner has more orchestral and instrumental composers such as {\em Holst, Elgar}, and {\em Stravinski}.
Not surprisingly, {\em Mozart} and {\em Beethoven} are the most popular composers in the classical music cluster. The islands of {\em Erik Satie} and {\em Arvo P{\"a}rt} connect the big island in the east with the west-most island of contemporary classical music represented by minimalists Philip Glass and Michael Nyman.

\section{Visualizing 
Recommendations}

\begin{figure*}[htb]
\begin{center}
\vspace{-.3cm}
\includegraphics*[width=18cm, angle = 0]{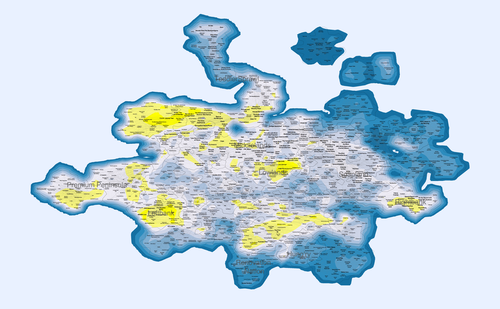}
\vspace{-.3cm}
\caption{\small\sf Personalized recommendation heat map for a typical but fictious user. Regions where the highest recommended shows sit are colored yellow. Regions of low scores are colored with dark cool colors. \label{heatMap}}
\end{center}
\end{figure*}

Once a global map of the item-item similarity is available, a personalized recommendation ``heat map'' 
can be generated, where regions of low interest are colored with cool colors, and regions containing highly recommended shows are colored with a hot color. Figure~\ref{heatMap} shows such a heat map, where 
shows are scored using a factorization based recommender~\cite{Hu_Koren_Volinsky_2008}, with blue color for 
shows that score low, and yellow for shows that score high. We generate such maps dynamically based on the viewing preference of the user, and based on what TV shows are available at this moment in time, much like a personalized weather forecast, but for TV shows. These maps uniquely capture the viewing preferences of the user/household, and evolve as the availability of TV shows, and as the user's taste, change with time. We can also generate a heat map profile, determined by how often the user watches certain shows over a fixed time period, say, a week or a month.

\begin{figure}[htb]
\begin{center}
\includegraphics*[width=8cm, angle = 0]{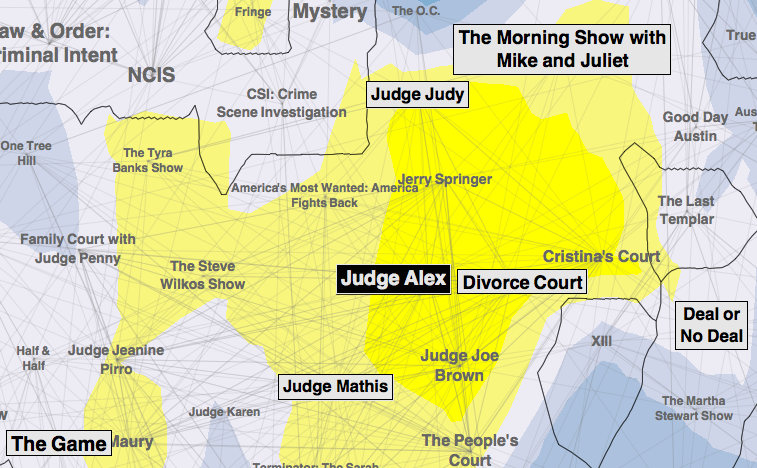}
\vspace{-.3cm}
\caption{\small\sf Explaining a recommendation.
Framed labels with a light background are shows the user watched, with font size proportional to the amount of time spent. The recommended show is framed with a black background. \label{judgeAlex}}
\end{center}
\end{figure}

Using the map interface, recommendations can be made in several different ways.  In a
user-driven mode, the heat map is presented to the user for exploration.
The user sees the areas of concentration of recommended shows or their viewing
behavior, and can explore the map to find other shows in the same
countries, or in neighboring countries.  In some cases, a user might
want to explore faraway lands for something new, or complete a
cross-country trip over time.  In a user-passive mode, recommendations
are made by the recommender system and are highlighted on the map,
along with the shows that were most similar.
In this mode, the map allows the user to understand why the recommendations are made. 
For example, Figure~\ref{judgeAlex} shows that {\em Judge Alex} is recommended because the user watched
{\em Divorce Court}, {\em Judge Judy}, and {\em Judge Mathis}. 
In addition to the recommended show, the map shows other nearby shows that are related and which might be worth exploring.

Note that the ``hot'' areas on the map in Figure~\ref{heatMap},
include news shows and kids shows. This could be an indication that
there are children and adults that household. 
Our TV data is on the household
level, and a household may contain various individuals with widely
varying preferences. The visualization of household viewing behavior
on a map could possibly separate out different elements of the
household and allow each member to get more personalized
recommendations, e.g., the 8-year-old boy will look for interesting
shows in {\em ToddlerSprawl}, while the teenagers look in the {\em
Premium Peninsula} and the parents in {\em Newsistan}.

\begin{figure}[htb]
\begin{center}
\includegraphics*[width=7.3cm, angle = 0]{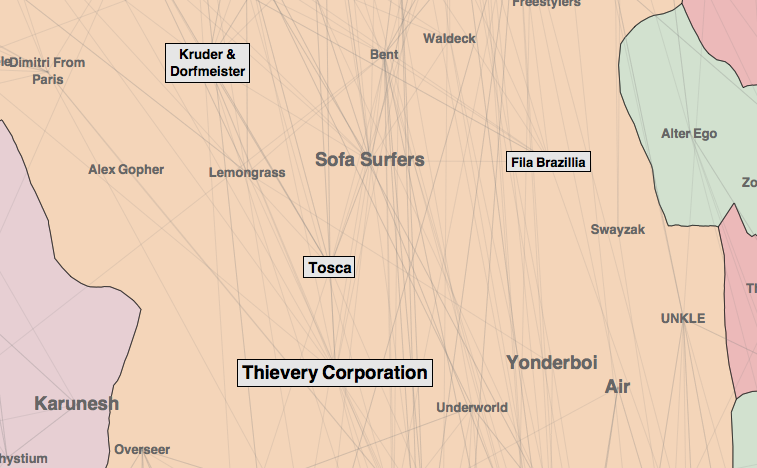}
\vspace{-.3cm}
\caption{\small\sf The neighborhood of music taste: framed labels indicate music the user has requested, with font size proportional to the number of songs played. \label{lastfmNeighborhood}}
\end{center}
\end{figure}

While {\tt last.fm} does not explicitly make recommendations, it does so implicitly, by playing music that is considered related to that often requested by the user. For example, one of the authors of this paper is a fan of the music groups {\em Thievery Corporation, Tosca, Kruder \& Dorfmeister, and Fila Brazillia}.  After several months of frequently requesting that last.fm play music similar to the above four groups, a look at the list of music most often played reveals that all  bands in the immediate vicinity (such as {\em Sofa Surfers, Air, Waldeck}) were played with high frequency; see Fig.~\ref{lastfmNeighborhood}. 

We envisage our maps as a navigational interface to interactive media services such as video/movie/music-on-demand, in which each title is a live link to the video/audio recording, and the user can zoom in and out to explore the land of TV/movies/music, much as online maps are used, except that here clicking on each ``town'' allows instant exploration.

\section{Related Work}

There is little previous work on generating map representations of graphs. 
Most related work deals with accurately and appealingly representing 
a given geographic region, or on re-drawing an 
existing map subject to additional constraints. Examples of 
the first kind of problem are found in traditional cartography, e.g., the 
Mercator's 1569 projection of the sphere into 2D Euclidean space.
Examples of the second kind of problem are found in cartograms, where the goal is to redraw a map so that the 
country areas are proportional to some metric, an idea which dates back to 1934~\cite{Raisz34} and is
still popular today (e.g., the New York Times red-blue maps of the US, showing the presidential election results in 2000 and 2004 with states drawn proportional to population). 

Somewhat similar to cartograms, treemaps~\cite{treemaps92}, squarified treemaps~\cite{squarified}, and the more recent newsmaps~\cite{newsmap} 
represent hierarchical information by means of space-filling tilings, allocating area proportional to some important metric. The map of New Yorkistan~\cite{newyorkistan} takes a real place, New York City, and creatively renames its neighborhoods. 
The map of science~\cite{mapofscience} provides an overview of the scientific landscape, based on citations of journal articles.

Placing imagined places on a map as if they were real countries also has a long history, 
e.g., the 1930's Map of Middle Earth by Tolkien~\cite{tolkien}. 
A more recent example is xkcd's
Map of Online Communities~\cite{xkcd}. 
While most such maps are 
generated in an ad hoc manner and are not 
strictly based on underlying data, they are often visually 
appealing.

\section{Conclusion and Future Work}

In this paper, we described a technique for visualizing relational datasets along with clustering information in the form of maps. This paradigm allows one to clearly see clusters and to identify patterns and trends. We believe that such maps can be a powerful tool for the visualization and explanation of recommendations, as well as for interactive navigation of media systems.

It is clear that there are practical as well theoretical obstacles to
obtaining ``perfect'' maps, maps that do not omit or distort
the underlying information. However, a similar drawback can be found
in any 2-dimensional representation of high-dimensional data. There are many interesting questions regarding the aesthetics of generated maps: What shapes would be best and why? Are low complexity convex shapes preferable? How do we create realistic borders? How do we deal with oceans, lakes, and rivers? While we have made a first step towards answering these questions, there is a great deal more that could be done.

It would also be worthwhile to look at the historical record and visualize the evolution of the TV landscape over time, looking at the impact of important TV events (e.g., presidential elections, American Idol finale) on
the TVLand. 

\bibliographystyle{abbrv}
\small{
\bibliography{paper}
}
\end{document}